\documentclass[12pt,titlepage]{article}
\usepackage{amsmath}
\usepackage{amssymb}
\usepackage{graphicx}
\usepackage{caption2}
\usepackage{amsfonts}
\usepackage{cite}
\usepackage{lineno}

\newcommand{\be}{\begin{equation}}
\newcommand{\ee}{\end{equation}}
\newcommand{\bea}{\begin{eqnarray}}
\newcommand{\eea}{\end{eqnarray}}
\newcommand{\bef}{\begin{figure}}
\newcommand{\ef}{\end{figure}}
\newcommand{\bt}{\begin{tabular}}
\newcommand{\et}{\end{tabular}}
\newcommand{\bno}{\begin{enumerate}}
\newcommand{\eno}{\end{enumerate}}

\def\3{\ss}

\catcode`\"=\active
\def"{\accent'177}

\begin{document}

\begin{center}

{\Large Influence of round-off errors on the reliability of  numerical simulations of chaotic dynamic systems}

\vspace{0.3cm}
Shijie Qin $^1$ and Shijun Liao $^{1,2}$
\vspace{0.3cm}

$^1$ School of Naval Architecture, Ocean and Civil Engineering,\\  Shanghai Jiaotong University, Shanghai 200240, China

$^2$ Ministry-of-Education Key Lab for Scientific and Engineering Computing, Shanghai 200240, China
 \end{center}

\hspace{-0.85cm} {\bf Abstract}  {\em We illustrate that, like the truncation error,  the round-off error has a significant influence on the reliability of numerical simulations of chaotic dynamic systems.  Due to the butterfly-effect, all numerical approaches in double precision cannot give a reliable simulation of chaotic dynamic systems.   So, in order to avoid man-made uncertainty of numerical simulations of chaos, we had to greatly decrease both of the truncation and round-off error to a small enough level, plus a verification of solution reliability  by means of  an additional computation using even smaller truncation and round-off errors.  }  

\section{Introduction} 

The sensitive dependence on initial conditions (SDIC) was first found in 1890  by Henri Poincar{\' e} \cite{Poincare1890} in a particular case of the three-body problem, who later proposed that such phenomena could be common, say in meteorology \cite{Wolfram2002}.  In 1963, using a digit computer (Royal McBee LGP-30) to solve a set of coupled ordinary differential equations (ODEs)
\begin{equation}
\left\{  
\begin{array}{lcl}
\dot x &= & - \sigma x+\sigma y,  \\
\dot y &= & r x- y-x\; z,  \\
\dot z &=&  x \; y-b z, 
\end{array} 
\right. \label{geq:Lorenz}
\end{equation}
where $\sigma, b$ and $0< r <+\infty$ are physical parameters,  Edward N. Lorenz \cite{Lorenz1963}  found  the so-called butterfly-effect: a tiny change in initial condition might result in large difference in a later state.   The tiny difference in initial condition of such kind of chaotic dynamic systems enlarges exponentially \cite{sprott2010}, which can be characterized by a positive Lyapunov exponent $\lambda$.    In other words, the maximum Lyapunov exponent of a chaotic dynamic system must be positive.   However,  Lorenz \cite{Lorenz2006} also reported that, by means of the Runge-Kutta method with data in double precision,   the maximum  Lyapunov exponents of numerical simulations of a chaotic dynamic system given by different values of time-step may  fluctuate  around  zero,  say, its value  constantly  changes  between  positive  and  negative  ones,  even if the initial condition is exactly the same and the time step becomes rather small.  Thus, the computer-generated numerical simulations of chaotic dynamic systems are sensitive not only to initial condition but also to numerical algorithms.   This is easy to understand, since there always exist the truncation and roundoff errors at any steps of numerical simulations of chaos,  which enlarge exponentially due to the so-called butterfly-effect.   In addition,  Teixeira et al.  \cite{Teixeira2007}  investigated  the time-step sensitivity of nonlinear  atmospheric models and  found that ``different time steps may lead to different model climates and even different regimes'',  thus  ``for chaotic systems, numerical convergence {\em cannot} be guaranteed {\em forever} ''.    

Hoover  et al. \cite{Hoover2015}  illustrated the Lyapunov's instability by comparing numerical simulations of a chaotic Hamiltonian system given by two Runge-Kutta  and  five symplectic integrators \cite{Yoshida1990, Farres2013, McLachlan2014} in {\em double precision}, and found that ``{\em all} numerical methods are susceptible to Lyapunov instability, which severely limits the maximum time for which chaotic solutions can be {\em accurate}'', although ``all of these  integrators conserve energy almost {\em perfectly}'' and ``they also reverse back to the initial conditions even when their trajectories are {\em inaccurate}''.  As reported by Hoover  et al. \cite{Hoover2015} , ``the {\em advantages} of higher-order methods are lost rapidly for typical chaotic Hamiltonians'', and ``there is little distinction between the symplectic and the Runge-Kutta integrators for chaotic problems, because both types lose accuracy  at the very same rate, determined by the maximum Lyapunov exponent.'' 

Even for some dynamic systems without Lyapunov's instability, it is rather hard to gain accurate prediction, too.   For example, let us consider the famous Lorenz equation (\ref{geq:Lorenz}) in the case of $\sigma=10, b=8/3$, which is chaotic only when $r\geq 24.74$.  It is well-known that, when $1<r<24.74$, the long-term solution of the Lorenz equations should finally tend to one of the two {\em stable} fixed points \[ C(\sqrt{b(r-1)},~\sqrt{b(r-1)},~r-1)\] and \[C'(-\sqrt{b(r-1)},~-\sqrt{b(r-1)},~r-1).\]    However, in the  case of $r=22$,  Li et al. \cite{JP-Li2001} studied the sensitive dependance of the  fixed point on the time-step $\Delta t$, which are calculated by means of many explicit/implicit  numerical approaches (such as Euler's method, Runge-Kutta methods of orders from 2 to 6, Taylor series methods of orders from 2 to 10, Adams methods of orders from 2 to 6, and so on) in {\em double precision}, but found that  the  long-term results of numerical simulations  are rather sensitive to the step size $\Delta t$, say, they always fluctuate between the two fixed points, no matter how small the time-step $\Delta t$ is.  Thus, they made the conclusion that ``numerical solution obtained by any stepsize is {\em unrelated} to exact solution'' \cite{JP-Li2001}.

These numerical  facts   lead  to some intense arguments.  Some even believed that ``all chaotic responses are simply {\em numerical noise} and have {\em nothing} to do with the solutions of differential equations'' \cite{Yao2008}.    On the other side,  using double precision data and a few examples based on the 15th-order Taylor-series procedure with decreasing time-step, Lorenz \cite{Lorenz2008} was optimistic and believed that ``numerical approximations can converge to a chaotic {\em true} solution throughout {\em any} finite range of time, although, if the range is large, confirming the convergence can be utterly {\em impractical}.'' 

Is it possible to gain a convergent solution of a chaotic dynamic system in a long enough interval of time?  This  question is of critical importance.         

Note that convergent chaotic simulations cannot be guaranteed even if different high-order numerical methods were used, as illustrated by many researchers \cite{Teixeira2007, Hoover2015, JP-Li2001}.   So, it is useless to reduce truncation errors only.      Note that all of  them used data in double precision, which leads to round-off error at each step,   which also enlarges exponentially due to the butterfly-effect of chaos, just like truncation errors.    So, to guarantee the convergence of chaotic solution, both of the truncation and round-off errors {\em must} be controlled to be much smaller than physical variables under investigation.   In 2009, Liao \cite{Liao2009} suggested the so-called ``Clean Numerical Simulation'' (CNS) \cite{Liao2013A, Liao2013B}  for chaotic dynamic systems and turbulence, which is based on the {\em arbitrary} order of Taylor expansion method \cite{Corliss1982, Barrio2005} and the use of all data in {\em arbitrary} precision (i.e. multiple precision \cite{Oyanarte1990}), plus a {\em verification} of solution reliability.   By means of the CNS using the 3500th-order Taylor expansion method and data in 4180-digit precision, the convergent, reliable chaotic solutions of Lorenz equation were obtained even in $[0,10000]$, a rather long interval of time \cite{Liao-Wang2014}.  Its solution reliability  was further verified by means of the CNS using the 3600th-order Taylor series method and data in 4515-digit precision \cite{Liao-Wang2014}.   This work supports Lorenz's  optimistic viewpoint that  ``numerical approximations can converge to a chaotic {\em true} solution throughout {\em any} finite range of time''  \cite{Lorenz2008}.   

Here, we further illustrate the effects of round-off error on the reliability of  numerical simulations of chaotic dynamic systems, and show the importance of reducing both of truncation and round-off errors.                                  
 
\section{Influence of round-off errors}

To investigate the  influence of round-off error on the reliability of numerical simulations of chaotic dynamic systems, let us consider the Lorenz equation (\ref{geq:Lorenz}) in the case of $\sigma=10, b=8/3$ and $r=23$, with the exact initial condition
\begin{equation}    
x(0)=5, \;\; y(0)=5, \;\;  z(0)=10.  \label{ic:Lorenz}
\end{equation} 
Since its solution is chaotic when $r\geq 24.74$,  the long-term numerical simulation should finally tend to one of the two stable fixed points in he case of $r=23$.   Let $h=\Delta t$ denote the step-size.   It is found that the final values of the numerical simulations given by the 4th-order Runge-Kutta method using data in double precision indeed rather sensitive to the step-size $h$, as shown in Fig.~1.    No matter how small the stepzise $h$, the final values always fluctuate between the two fixed points.  The same phenomenon was reported by Li et al. \cite{JP-Li2001}.      

\begin{figure}[t]
    \begin{center}
        \begin{tabular}{cc}   
            \includegraphics[width=4in]{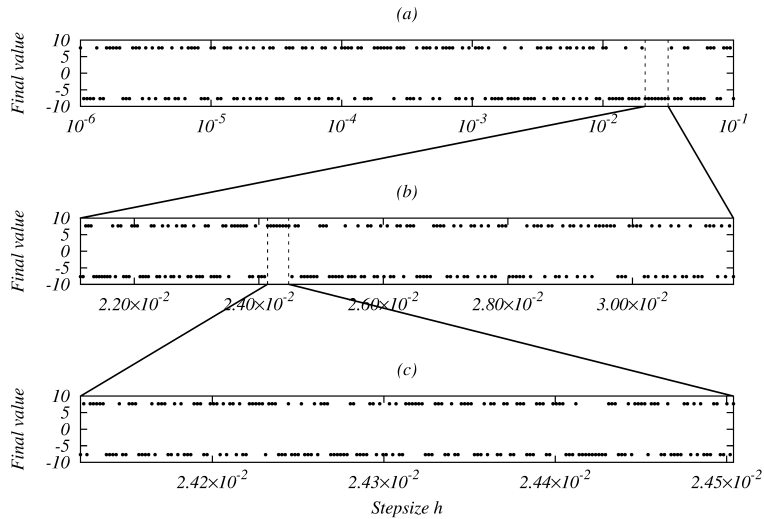} 
        \end{tabular}
    \caption{Final value of \emph{x(t)} of the Lorenz equations in case of $r=23$ with the initial value (5,5,10) versus \emph{stepsize h},  obtained by the 4th-order Runge-Kutta method in double precision. The \emph{stepsize h} varies from (a) $10^{-6}$ to $10^{-1}$, (b) from $2.11349 \times 10^{-2}$ to $3.16228 \times 10^{-2}$ and (c) from $2.41107 \times 10^{-2}$ to $2.45086 \times 10^{-2}$, respectively.}
    \end{center}
    \label{Fig:stepsize-sensitivity:RK}
\end{figure}

In order to reduce the truncation error, we use the $M$th-order Taylor expansion method and the stepsize $h=0.01$ to gain numerical simulations of Lorenz equation (\ref{geq:Lorenz}) in the case of $\sigma=10, b=8/3$ and $r=23$ with the initial condition (\ref{ic:Lorenz}).    Obviously, as $M$ becomes large, the truncation error could be rather small.   To increase the computation efficiency, we use parallel computation by means of different numbers of processes, denoted by $np$.  

It is found that, using the double precision, we gain different final values by means of high-order Taylor expansion method even using the {\em same} laptop (Thinkpad L440 with Intel Core i7-4712MQ) but {\em different} numbers of processes, as shown in Fig.~2.   Even if the order  of Taylor expansion method is rather high, such as $M=200$, corresponding to rather small truncation error, we still gain different final values using the same laptop but different numbers of processes, as shown in Fig.~3.   It is found that, using the high-order Taylor series method in double precision, we would gain different final values using the {\em same} number of processes $np$ but {\em different} computers,  as shown in Fig.~4.  This kind of man-made uncertainty of numerical simulations {\em cannot} be avoided even by means of rather high order of Taylor series method such as $M=200$, as shown in Fig.~5.     All of these  illustrate that decreasing truncation error {\em alone} cannot avoid the man-made uncertainty of numerical simulations for the considered problem.            

As pointed out by Monniaux \cite{Monniaux2008},  even using the {\em same} programmes with the {\em same} compiler,  which have exactly  the  {\em same} expression, the {\em same} values in the {\em same} variables and so on,  different working platforms may exhibit subtle differences with respect to floating-point computations.  Thus, both of the different number of processes operating on the same computer and the different computers with the same number of processes  can  generate  rather  tiny  difference of  round-off errors, which unfortunately would be enlarged so greatly (due to the butterfly-effect of chaos) that {\em completely} different numerical simulations might be obtained!   
In practice, round-off error sometimes indeed might lead to some serious problems, such as the system failure in the military: on February 25, 1991, a loss of significance in a MIM-104 Patriot missile battery prevented it from intercepting an incoming Scud missile in Dhahran, Saudi Arabia, contributing to the death of 28 soldiers from the U.S. Army's 14th Quartermaster Detachment \cite{Office1992Patriot}.    

Such kind of man-made uncertainty of numerical simulations of the Lorenz equation can be avoided by  decreasing  both of the truncation and round-off errors at the same time!   It is found that, using the $M$th-order Taylor series method ($M\geq 130$)  in the 512-digit precision,  {\em all}  numerical simulations agree quite well in the {\em whole} interval of time and besides tend to the {\em same} fixed point, even if {\em different} numbers of processes are used on {\em different} computers, as shown in Figs.~6-8.    All of these examples illustrate the importance of decreasing both of truncation and round-off errors to the reliability of numerical simulations of chaotic dynamic systems, such as Lorenz equation, three-body problem and so on.   Therefore, it clearly indicates that, using numerical approaches in double precision, one can not avoid the man-made uncertainty of numerical simulations for chaotic dynamic systems, as shown by many researchers \cite{Lorenz1963,  Lorenz2006, Teixeira2007, Yao2008, JP-Li2001,Hoover2015}.    

The above-mentioned  examples also explain why one had to use the 3500th-order Taylor expansion method in the 4180-digit precision \cite{Liao-Wang2014}  so as to gain a convergent numerical simulation of a chaotic solution of Lorenz equation in a rather long interval of time [0,10000], whose reliability was further verified by means of the 3600th-order Taylor series method and data in 4515-digit precision.   Nowadays,  the importance of using multiple-precision  data  to  gain  reliable  numerical  simulations  of  chaotic dynamic systems  receives  recognition  by  more and more  researchers    \cite{Logg2010AIP, Sarra2011, Wang2012, Wang2016, Barrio2015, Liao2016CNS}.  

\section{Concluding remarks and discussions}

We confirm that, due to the butterfly effect,  the traditional numerical approaches in double precisions indeed cannot give reliable numerical simulations of chaotic dynamic systems.   Thus,   decreasing the truncation error {\em alone} cannot avoid the man-made uncertainty of numerical simulations of chaos.    However,  such kind of man-made uncertainty of numerical simulations for chaotic dynamic systems can be {\em completely} avoided by  decreasing {\em both} of the truncation and round-off errors at the same time, plus a {\em verification} of solution reliability by means of additional computations using even smaller truncation and round-off errors.    

In this paper we illustrate that, due to the butterfly-effect,  even the very tiny difference of round-off error caused by different numbers of processes or different  computers might lead to significant variation of numerical simulations of a chaotic dynamic system.  So,   the butterfly-effect is indeed a huge obstruction for us  to  gain reliable numerical simulations of chaos in a long interval of time.  Note that a few current numerical investigations suggest that turbulent flows might be sensitive even to micro-level thermal fluctuation \cite{Liao2016CNS}.  Then naturally,  the turbulent flows should be also sensitive to numerical noises.  Thus,  it should be of benefit to study the influence of numerical noises to numerical simulations of turbulence

\begin{figure}[t]
    \begin{center}
        \begin{tabular}{cc}
            \includegraphics[width=2.8in]{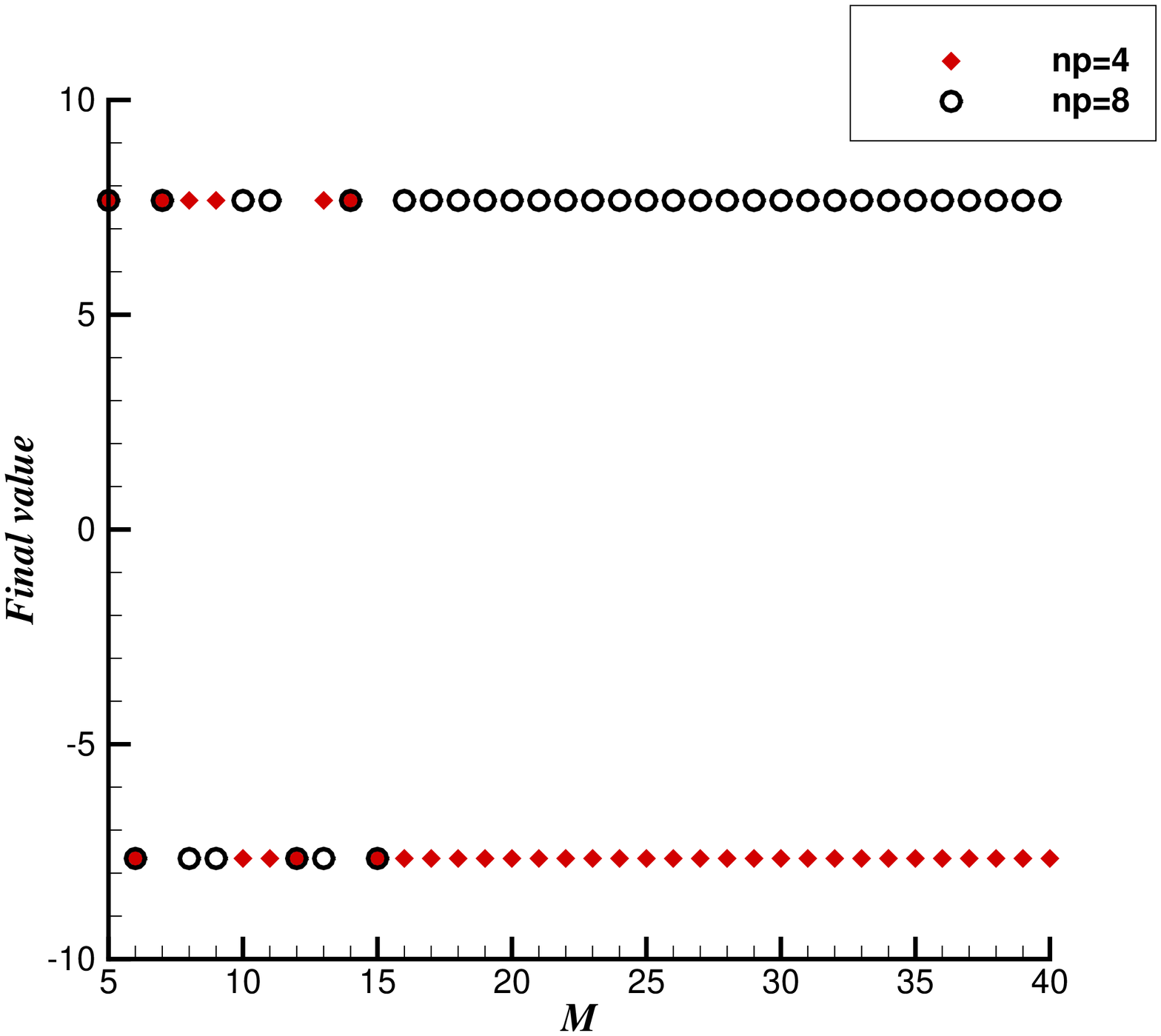} \\
        \end{tabular}
    \caption{Final values of numerical simulation of $x(t)$ of the Lorenz equations in case of $r=23$ with the initial value (5,5,10) versus $M$ (i.e. the truncated $M$th-order Taylor's expansion method),  given by the same laptop (Thinkpad L440 with Intel Core i7-4712MQ) using data in double-precision but the different \emph{np} (number of processes).}
    \end{center}
    \label{Fig:sensitive-np:M=40}

    \begin{center}
        \begin{tabular}{cc}
            \includegraphics[width=2.8in]{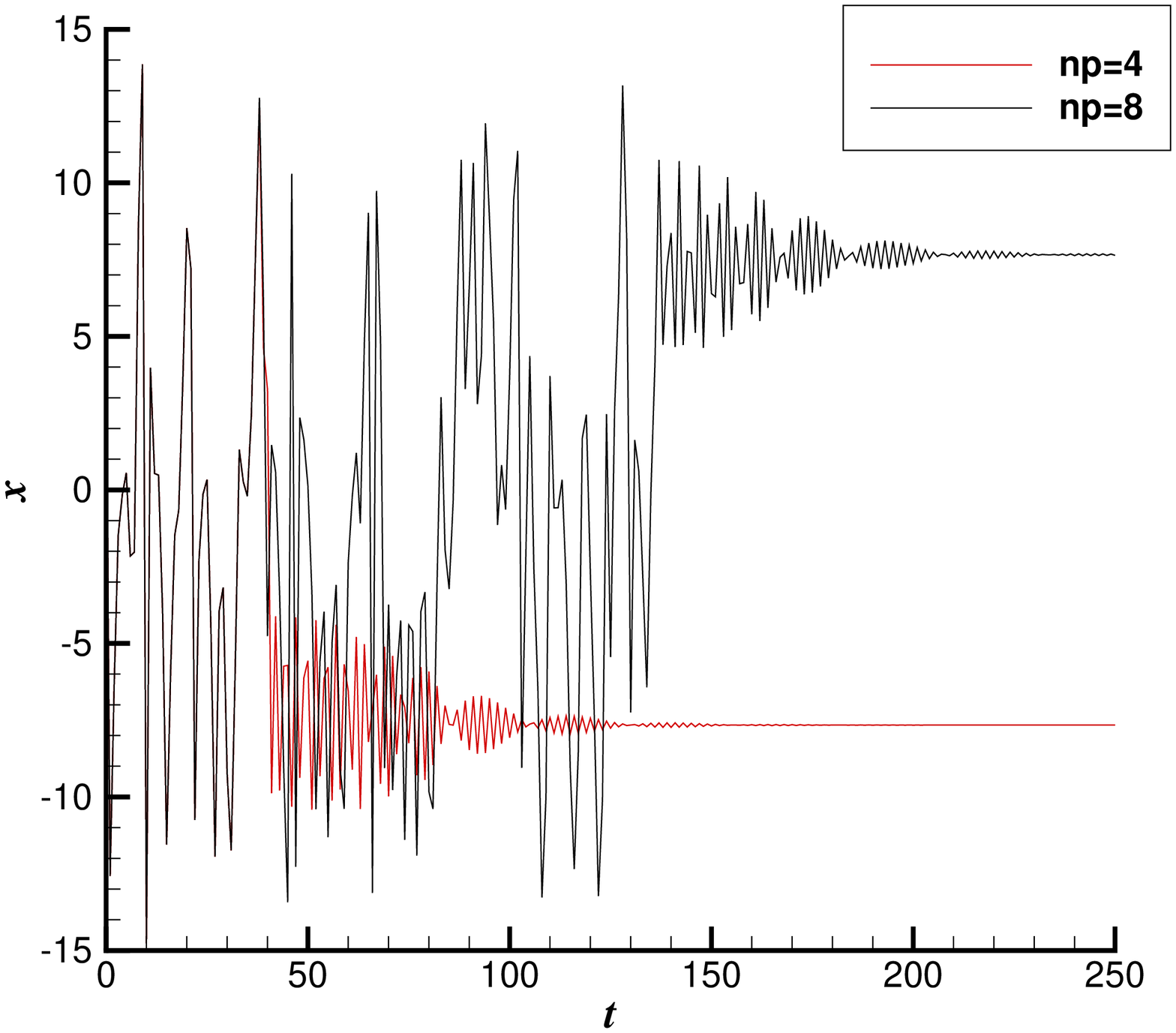} \\
        \end{tabular}
    \caption{The numerical simulations of $x(t)$ of the Lorenz equations in case of $r=23$ with the initial value (5,5,10),   given by the same laptop (Thinkpad L440 with Intel Core i7-4712MQ) using the 200th-order Taylor' expansion method (i.e. $M=200$) and data in double precision but the different \emph{np} (number of processes).}
    \end{center}
    \label{Fig:sensitive-np:M=200}
\end{figure}

\begin{figure}[t]
    \begin{center}
        \begin{tabular}{cc}
            \includegraphics[width=2.8in]{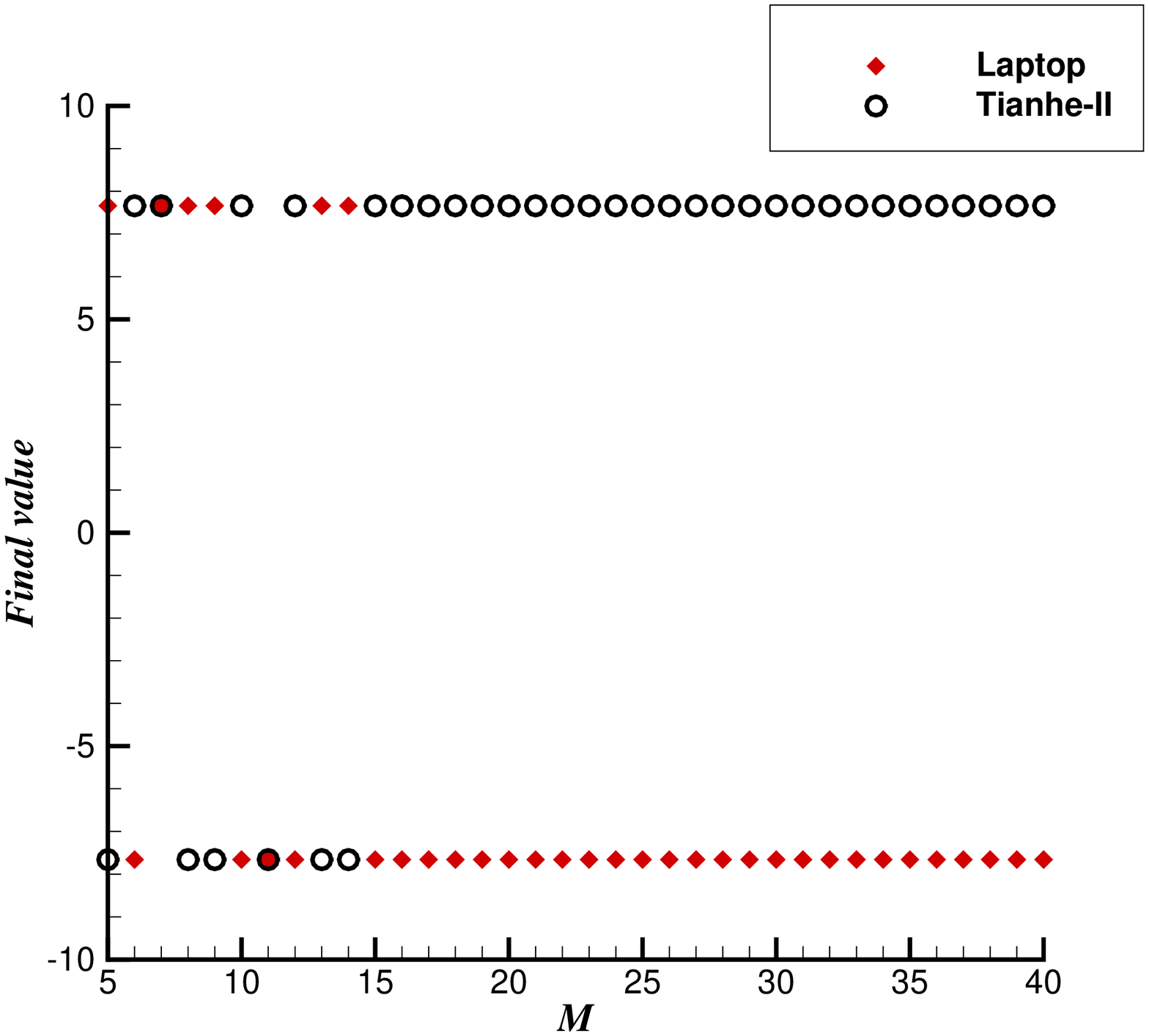} \\
        \end{tabular}
    \caption{Final value of numerical simulations of $x(t)$ versus \emph{M} (i.e., the truncated $M$th-order Taylor's expansion),  given by means of the same number of processes (\emph{np}=4) and data in double precision but the different computers, i.e.  the laptop (Thinkpad L440 with Intel Core i7-4712MQ) and the supercomputer Tianhe-II, respectively.}
    \end{center}

    \begin{center}
        \begin{tabular}{cc}
            \includegraphics[width=2.8in]{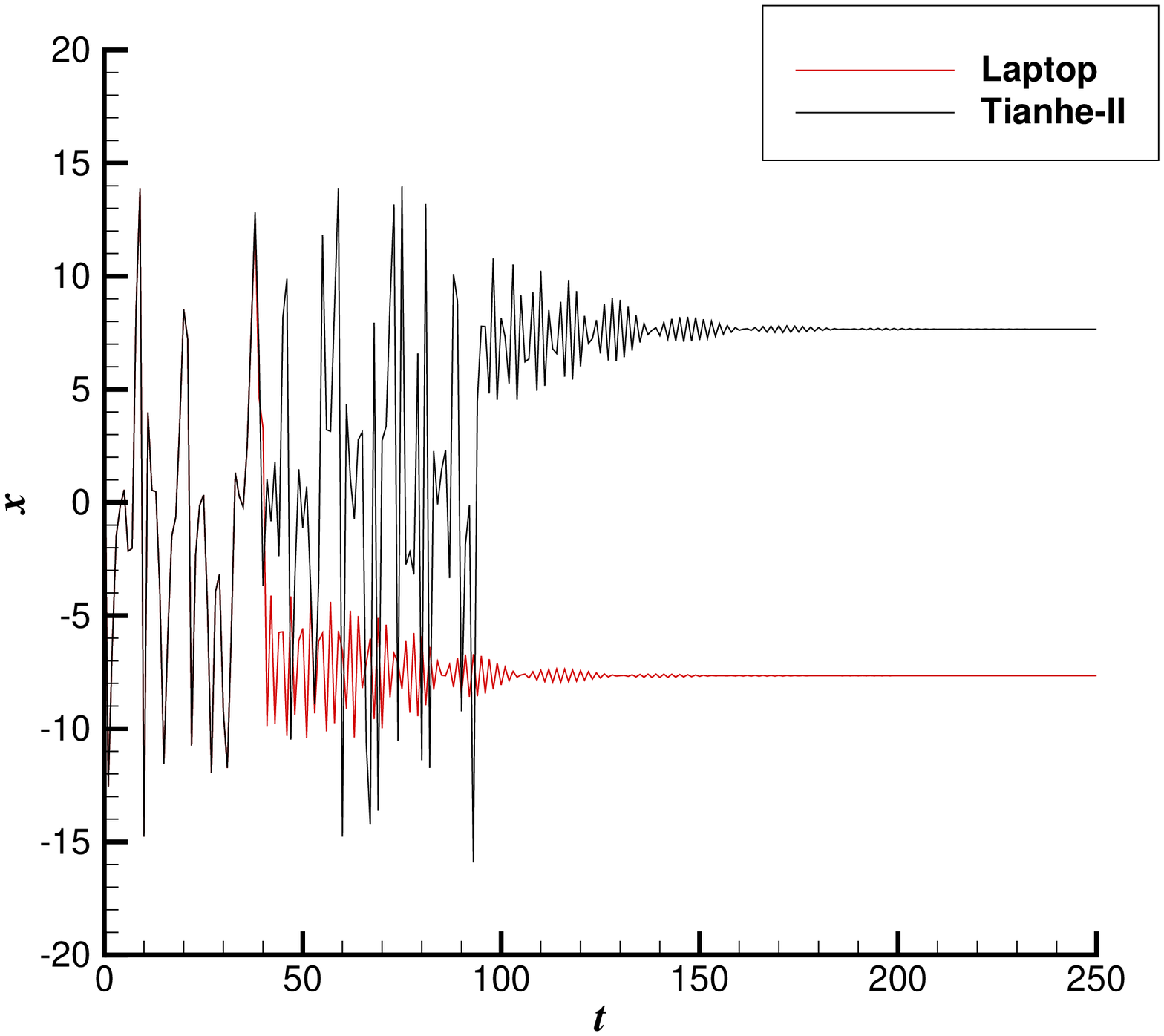} \\
        \end{tabular}
    \caption{The numerical simulations of $x(t)$, given by the 200th-order Taylor's expansion method  (i.e. $M=200$) using the same number of processes (\emph{np}=4)  and data in double precision but the different computers, i.e. the laptop (Thinkpad L440 with Intel Core i7-4712MQ) and the supercomputer Tianhe-II, respectively.}
    \end{center}
\end{figure}

\begin{figure}[t]
    \begin{center}
        \begin{tabular}{cc}
            \includegraphics[width=4in]{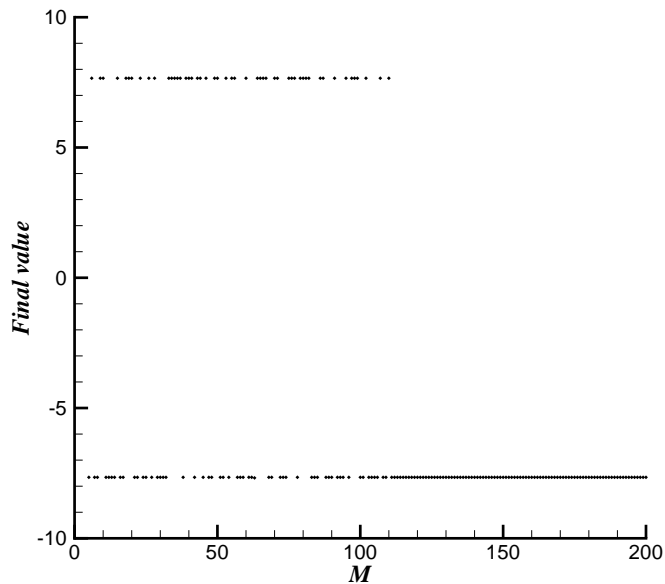} \\
        \end{tabular}
    \caption{ Final value of the numerical simulation of $x(t)$ of the Lorenz equations in case of $r=23$ with the initial value (5,5,10) given by means of the $M$th-order Taylor's expansion using data in 512-digit  precision.  For all given $M$, the same final values of $x(t)$ are obtained, which are independent of the computers (i.e. the laptop and the supercomputer) and the number of precesses ($np=4$ or 8).  As $M\geq 130$, the final value becomes to be independent of $M$, too.}
    \end{center}
\end{figure}

\begin{figure}
    \begin{center}
        \begin{tabular}{cc}
            \includegraphics[width=3in]{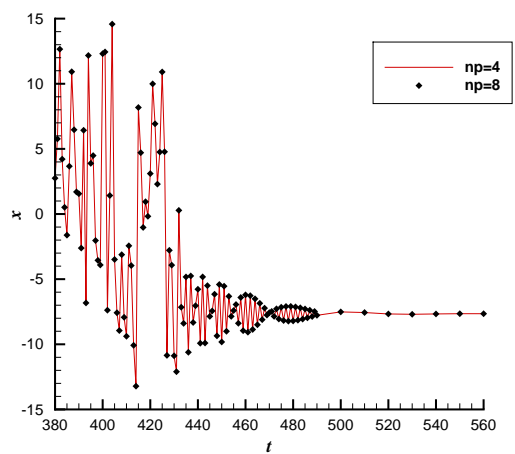} \\
        \end{tabular}
    \caption{The numerical simulations of $x(t)$  given by the 200th-order Taylor's expansion method (i.e. $M=200$) and data in 512-digit  precision using the same laptop (Thinkpad L440 with Intel Core i7-4712MQ) but different \emph{np} (number of processes).}
    \end{center}
\end{figure}

\begin{figure}
    \begin{center}
        \begin{tabular}{cc}
            \includegraphics[width=3in]{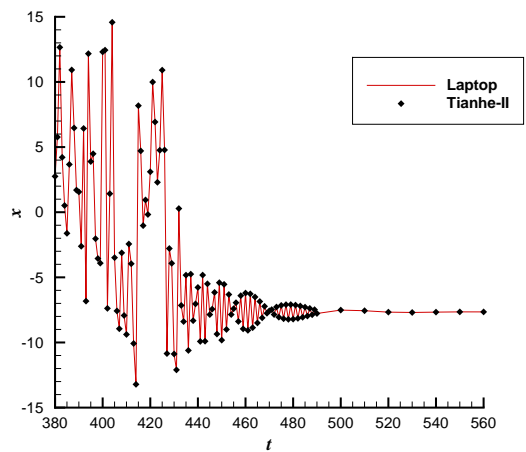} \\
        \end{tabular}
    \caption{The numerical simulations of $x(t)$ given by means of the 200th-order Taylor's expansion method (i.e. $M=200$) and data in 512-digit precision  using the same number of processes ($np$=4) but  the different computers, i.e. the laptop (Thinkpad L440 with Intel Core i7-4712MQ) and the supercomputer Tianhe-II, respectively. }
    \end{center}
\end{figure}

\bibliographystyle{elsarticle-num}
\bibliography{chaos}

\end{document}